\documentstyle[editedbk,epic]{kluwerbk}
\newcommand{\BEQ}{\begin{equation} }
\newcommand{\EEQ}{\end{equation} }
\newcommand{\BEA}{\begin{eqnarray} }
\newcommand{\EEA}{\end{eqnarray} }
\newcommand{\BEAn}{\begin{eqnarray*} }
\newcommand{\EEAn}{\end{eqnarray*} }
\newcommand{\Balk}[1]{\rule[-#1em]{0em}{#1em}}
\newcommand{\oo}{\mbox{$\frac{1}{N}$} }
\newcommand{\ON}{\mbox{O($N$)}}
\newcommand{\EWL}{\Big\langle}
\newcommand{\EWR}{\Big\rangle}
\newcommand{\OO}{\mbox{${\cal O}(\frac{1}{N})$ }}
\newcommand{\rf}[1]{\mbox{(\ref{#1})}}
\newcommand{\al}{\mbox{$\alpha$}}
\newcommand{\vs}{\mbox{$\vec{S}$}}
\newcommand{\ds}{\displaystyle}
\newcommand{\qp}{\mbox{qp - field }}
\newcommand{\qps}{\mbox{qp - fields }}

\newcommand{\Theo}[1]{
\begin{list}{{\it Theorem}
\arabic{theo}:}{\labelwidth6em\leftmargin6em\labelsep1em}
\item #1
\end{list}\refstepcounter{theo}}

\newcommand{\Caption}[2]{
\refstepcounter{figure}
\begin{center}
Figure \arabic{figure}: #2
\end{center}
\addcontentsline{lof}{figure}{#1}
\addtocontents{lof}{#1}}

\newcommand{\YoungS}[1]{\unitlength0.5em \begin{picture}(#1,1)
 \multiput(0,0)(1,0){#1}{\line(0,1){1}} \multiput(0,0)(0,1){2}{\line(1,0){#1}}
 \put(#1,0){\line(0,1){1}}
 \end{picture} }

\newcommand{\YoungSg}[1] {\unitlength0.5em \begin{picture}(8,1)
 \multiput(0,0)(1,0){3}{\line(0,1){1}} \multiput(0,0)(0,1){2}{\line(1,0){6}}
 \put(6,0){\line(0,1){1}}
 \put(6.1,0){\makebox(0,0)[l]{$\:{\scriptscriptstyle #1 }\; $ }}
 \end{picture} }

\newcommand{\YoungA}[1]{\unitlength0.25em \begin{picture}(2, #1 )(0,-1)
 \multiput(0,0)(2,0){2}{\line(0,1){#1}}
 \multiput(0,0)(2,0){2}{\line(0,-1){#1}}
 \multiput(0,-#1)(0,2){#1}{\line(1,0){2}}
 \put(0,#1){\line(1,0){2}}
 \end{picture} }

\newcommand{\aStower}{\setlength{\unitlength}{0.012in}
\begin{picture}(180,175)(12,10)
\drawline(35,40)(40,40)
\drawline(35,60)(40,60)
\drawline(35,80)(40,80)
\drawline(35,100)(40,100)
\drawline(35,120)(40,120)
\drawline(35,140)(40,140)
\drawline(35,160)(40,160)
\drawline(55,20)(55,25)
\drawline(95,20)(95,25)
\drawline(135,20)(135,25)
\drawline(175,20)(175,25)
\thicklines
\drawline(45,40)(65,40)
\drawline(85,100)(105,100)
\drawline(85,120)(105,120)
\drawline(85,140)(105,140)
\drawline(85,160)(105,160)
\drawline(125,120)(145,120)
\drawline(125,140)(145,140)
\drawline(125,158)(145,158)\drawline(125,162)(145,162)
\drawline(165,160)(185,160)
\thinlines
\drawline(37.000,182.000)(35.000,190.000)(33.000,182.000)
\drawline(35,190)(35,20)(195,20)
\drawline(187.000,18.000)(195.000,20.000)(187.000,22.000)
\put(55,0){\makebox(0,0)[b]{\footnotesize 0}}
\put(95,0){\makebox(0,0)[b]{\footnotesize 2}}
\put(135,0){\makebox(0,0)[b]{\footnotesize 4}}
\put(175,0){\makebox(0,0)[b]{\footnotesize 6}}
\put(5,40){\makebox(0,0)[l]{\footnotesize $\mu-1$}}
\put(5,60){\makebox(0,0)[l]{\footnotesize $\mu$}}
\put(5,80){\makebox(0,0)[l]{\footnotesize $\mu+1$}}
\put(5,100){\makebox(0,0)[l]{\footnotesize $\mu+2$}}
\put(5,120){\makebox(0,0)[l]{\footnotesize $\mu+3$}}
\put(5,140){\makebox(0,0)[l]{\footnotesize $\mu+4$}}
\put(5,160){\makebox(0,0)[l]{\footnotesize $\mu+5$}}
\put(55,30){\makebox(0,0)[b]{\footnotesize $S$}}
\put(95,90){\makebox(0,0)[b]{\footnotesize $l=1$}}
\put(135,110){\makebox(0,0)[b]{\footnotesize $l=0$}}
\put(175,150){\makebox(0,0)[b]{\footnotesize $l=0$}}
\put(185,0){\makebox(0,0)[b]{\footnotesize $t$}}
\put(30,175){\makebox(0,0)[rb]{\footnotesize $[\delta]$}}
\end{picture} }

\newcommand{\standardtower}{\setlength{\unitlength}{0.012in}
\begin{picture}(185,175)(5,10)
\drawline(35,40)(40,40)
\drawline(35,60)(40,60)
\drawline(35,80)(40,80)
\drawline(35,100)(40,100)
\drawline(35,120)(40,120)
\drawline(35,140)(40,140)
\drawline(35,160)(40,160)
\drawline(55,20)(55,25)
\drawline(95,20)(95,25)
\drawline(135,20)(135,25)
\drawline(175,20)(175,25)
\thicklines
\drawline(45,40)(65,40)
\drawline(45,60)(65,60)
\drawline(45,80)(65,80)
\drawline(45,100)(65,100)
\drawline(45,120)(65,120)
\drawline(45,140)(65,140)
\drawline(45,160)(65,160)
\drawline(85,80)(105,80)
\drawline(85,100)(105,100)
\drawline(85,120)(105,120)
\drawline(85,140)(105,140)
\drawline(85,160)(105,160)
\drawline(125,120)(145,120)
\drawline(125,140)(145,140)
\drawline(125,160)(145,160)
\drawline(165,160)(185,160)
\thinlines
\drawline(37.000,182.000)(35.000,190.000)(33.000,182.000)
\drawline(35,190)(35,20)(195,20)
\drawline(187.000,18.000)(195.000,20.000)(187.000,22.000)
\put(55,0){\makebox(0,0)[b]{\footnotesize 0}}
\put(95,0){\makebox(0,0)[b]{\footnotesize 2}}
\put(135,0){\makebox(0,0)[b]{\footnotesize 4}}
\put(175,0){\makebox(0,0)[b]{\footnotesize 6}}
\put(-2 ,40){\makebox(0,0)[l]{\footnotesize $[\delta_0]$}}
\put(-2 ,60){\makebox(0,0)[l]{\footnotesize $[\delta_0]+1$}}
\put(-2 ,80){\makebox(0,0)[l]{\footnotesize $[\delta_0]+2$}}
\put(-2 ,100){\makebox(0,0)[l]{\footnotesize $[\delta_0]+3$}}
\put(-2 ,120){\makebox(0,0)[l]{\footnotesize $[\delta_0]+4$}}
\put(-2 ,140){\makebox(0,0)[l]{\footnotesize $[\delta_0]+5$}}
\put(-2 ,160){\makebox(0,0)[l]{\footnotesize $[\delta_0]+6$}}
\put(55,30){\makebox(0,0)[b]{\footnotesize $l_0$}}
\put(55,50){\makebox(0,0)[b]{\footnotesize $l_0+1$}}
\put(55,70){\makebox(0,0)[b]{\footnotesize $l_0+2$}}
\put(95,70){\makebox(0,0)[b]{\footnotesize $l_0$}}
\put(135,110){\makebox(0,0)[b]{\footnotesize $l_0$}}
\put(175,150){\makebox(0,0)[b]{\footnotesize $l_0$}}
\put(185,0){\makebox(0,0)[b]{\footnotesize $t$}}
\put(30,175){\makebox(0,0)[rb]{\footnotesize $[\delta]$}}
\end{picture} }

\newcommand{\atower}{\setlength{\unitlength}{0.012in}
\begin{picture}(180,175)(10,10)
\drawline(35,40)(40,40)
\drawline(35,60)(40,60)
\drawline(35,80)(40,80)
\drawline(35,100)(40,100)
\drawline(35,120)(40,120)
\drawline(35,140)(40,140)
\drawline(35,160)(40,160)
\drawline(55,20)(55,25)
\drawline(95,20)(95,25)
\drawline(135,20)(135,25)
\drawline(175,20)(175,25)
\thicklines
\drawline(45,40)(65,40)
\drawline(85,80)(105,80)
\drawline(85,120)(105,120)
\drawline(85,160)(105,160)
\drawline(125,120)(145,120)
\drawline(125,160)(145,160)
\drawline(165,160)(185,160)
\thinlines
\drawline(37.000,182.000)(35.000,190.000)(33.000,182.000)
\drawline(35,190)(35,20)(195,20)
\drawline(187.000,18.000)(195.000,20.000)(187.000,22.000)
\put(55,0){\makebox(0,0)[b]{\footnotesize 2}}
\put(95,0){\makebox(0,0)[b]{\footnotesize 4}}
\put(135,0){\makebox(0,0)[b]{\footnotesize 6}}
\put(175,0){\makebox(0,0)[b]{\footnotesize 8}}
\put(20,40){\makebox(0,0)[l]{\footnotesize $2$}}
\put(20,80){\makebox(0,0)[l]{\footnotesize $4$}}
\put(20,120){\makebox(0,0)[l]{\footnotesize $6$}}
\put(20,160){\makebox(0,0)[l]{\footnotesize $8$}}
\put(55,30){\makebox(0,0)[b]{\footnotesize $\alpha$}}
\put(185,0){\makebox(0,0)[b]{\footnotesize $t$}}
\put(30,175){\makebox(0,0)[rb]{\footnotesize $[\delta]$}}
\end{picture} }

\newcommand{\Ttower}{\setlength{\unitlength}{0.012in}
\begin{picture}(185,175)(5,10)
\drawline(35,40)(40,40)
\drawline(35,60)(40,60)
\drawline(35,80)(40,80)
\drawline(35,100)(40,100)
\drawline(35,120)(40,120)
\drawline(35,140)(40,140)
\drawline(35,160)(40,160)
\drawline(55,20)(55,25)
\drawline(95,20)(95,25)
\drawline(135,20)(135,25)
\drawline(175,20)(175,25)
\thicklines
\drawline(45,80)(65,80)
\drawline(45,120)(65,120)
\drawline(45,160)(65,160)
\drawline(85,80)(105,80)
\drawline(85,120)(105,120)
\drawline(85,140)(105,140)
\drawline(85,160)(105,160)
\drawline(125,120)(145,120)
\drawline(125,160)(145,160)
\drawline(165,160)(185,160)
\thinlines
\drawline(37.000,182.000)(35.000,190.000)(33.000,182.000)
\drawline(35,190)(35,20)(195,20)
\drawline(187.000,18.000)(195.000,20.000)(187.000,22.000)
\put(55,0){\makebox(0,0)[b]{\footnotesize 0}}
\put(95,0){\makebox(0,0)[b]{\footnotesize 2}}
\put(135,0){\makebox(0,0)[b]{\footnotesize 4}}
\put(175,0){\makebox(0,0)[b]{\footnotesize 6}}
\put(-2,40){\makebox(0,0)[l]{\footnotesize $2\mu-2$}}
\put(-2,80){\makebox(0,0)[l]{\footnotesize $2\mu$}}
\put(-2,120){\makebox(0,0)[l]{\footnotesize $2\mu+2$}}
\put(-2,160){\makebox(0,0)[l]{\footnotesize $2\mu+4$}}
\put(55,70){\makebox(0,0)[b]{\footnotesize $T$}}
\put(185,0){\makebox(0,0)[b]{\footnotesize $t$}}
\put(30,175){\makebox(0,0)[rb]{\footnotesize $[\delta]$}}
\end{picture} }

\newcommand{\Sprop}{
\setlength{\unitlength}{0.008in}
\begin{picture}(60,15)(-10,13)
\thicklines
\drawline(3,15)(37,15)
\end{picture} }

\newcommand{\SproprenO}{
\setlength{\unitlength}{0.008in}
\begin{picture}(100,15)(-30,13)
\thicklines
\drawline(-17,15)(-3,15)
\dashline[50]{6.000}(3,15)(37,15)
\drawline(43,15)(57,15)
\drawline (1.509,18.5)
        (2.768,21.674)
        (5.000,25.000)
\put(0,15){\circle*{6}}
\drawline(5,25) (9.769,28.025)
        (12.609,29.241)
        (15.000,30.000)
\drawline(15,30)        (17.515,30.105)
        (20.000,30.000)
\drawline(20,30)        (22.485,30.105)
        (25.000,30.000)
\drawline(25,30)        (27.391,29.241)
        (30.231,28.025)
        (35.000,25.000)
\drawline(35,25)        (37.232,21.674)
        (38.491,18.5)
\put(40,15){\circle*{6}}
\end{picture} }

\newcommand{\aprop}{
\setlength{\unitlength}{0.008in}
\begin{picture}(60,15)(-10,13)
\thicklines
\dashline[50]{6}(3,15)(37,15)
\end{picture} }

\newcommand{\aproprenA}{
\setlength{\unitlength}{0.008in}
\begin{picture}(100,15)(-30,13)
\thicklines
\dashline[50]{6}(-17,15)(-3,15)
\dashline[50]{6}(43,15)(57,15)
\drawline (1.509,18.5)
        (2.768,21.674)
        (5.000,25.000)
\put(0,15){\circle*{6}}
\drawline(5,25) (9.769,28.025)
        (12.609,29.241)
        (15.000,30.000)
\drawline(15,30)        (17.515,30.105)
        (20.000,30.000)
\drawline(20,30)        (22.485,30.105)
        (25.000,30.000)
\drawline(25,30)        (27.391,29.241)
        (30.231,28.025)
        (35.000,25.000)
\drawline(35,25)        (37.232,21.674)
        (38.491,18.5)
\drawline (1.509,11.5)
        (2.768,8.326)
        (5.000,5.000)
\drawline(5,5)  (9.769,1.975)
        (12.609,0.759)
        (15.000,0.000)
\drawline(15,0) (17.226,-0.313)
        (20.000,-0.418)
        (22.774,-0.313)
        (25.000,0.000)
\drawline(25,0) (27.391,0.759)
        (30.231,1.975)
        (35.000,5.000)
\drawline(35,5) (37.232,8.326)
        (38.491,11.5)
\put(40,15){\circle*{6}}
\end{picture} }

\newcommand{\Vertex}{\setlength{\unitlength}{0.006in}
\begin{picture}(50,55)(0,25)
\thicklines
\put(25,25){\circle*{10}}
\dashline[50]{10.000}(25,65)(25,30)
\drawline(20,20)(0,0)
\drawline(30,20)(50,0)
\end{picture} }

\newcommand{\VertexA}{\setlength{\unitlength}{0.005in}
\begin{picture}(160,115)(0,60)
\thicklines
\put(80,80){\circle*{10}}
\put(40,40){\circle*{10}}
\put(120,40){\circle*{10}}
\drawline(75,75)(45,45)
\drawline(35,35)(0,0)
\drawline(85,75)(115,45)
\drawline(125,35)(160,0)
\dashline[50]{10.000}(45,40)(115,40)
\dashline[50]{10.000}(80,160)(80,85)
\end{picture}}

\newcommand{\fusetree}{\setlength{\unitlength}{0.01in}
\begin{picture}(163,170)(0,10)
\thicklines
\drawline(62,0)(62,40)(32,60)(12,40)
\drawline(32,60)(12,80)
\drawline(62,40)(77,65)
\drawline(77,65)(107,80)(132,70)(142,45)
\drawline(132,70)(152,85)
\drawline(107,80)(107,110)
\drawline(77,65)(57,95)
\drawline(57,95)(37,110)
\drawline(37,110)(12,100)
\drawline(37,110)(17,135)
\drawline(57,95)(72,125)
\drawline(72,125)(57,145)
\drawline(72,125)(97,140)
\drawline(97,140)(97,170)
\drawline(97,140)(127,140)
\drawline(57,145)(57,170)
\drawline(57,145)(32,150)
\put(72,15){\makebox(0,0)[lb]{$\Phi$}}
\put(147,40){\makebox(0,0)[lb]{$S$}}
\put(157,80){\makebox(0,0)[lb]{$\alpha$}}
\put(107,115){\makebox(0,0)[b]{$S$}}
\put(97,175){\makebox(0,0)[b]{$\alpha$}}
\put(57,175){\makebox(0,0)[b]{$S$}}
\put(92,60){\makebox(0,0)[b]{$B$}}
\put(72,45){\makebox(0,0)[lb]{$C$}}
\put(7,35){\makebox(0,0)[rb]{$\alpha$}}
\put(7,75){\makebox(0,0)[rb]{$\alpha$}}
\put(7,95){\makebox(0,0)[rb]{$S$}}
\put(12,130){\makebox(0,0)[rb]{$\alpha$}}
\put(27,145){\makebox(0,0)[rb]{$S$}}
\put(62,70){\makebox(0,0)[rb]{$A$}}
\put(127,135){\makebox(0,0)[lb]{$S$}}
\end{picture} }

\newcommand{\amplitude}[5]{\mbox{
\setlength{\unitlength}{0.0125in}
\begin{picture}(206,50)(0,45)
\thicklines
\put(53,45){\circle{42}}
\put(153,45){\circle{40}}
\drawline(38,60)(13,85)
\drawline(38,30)(13,5)
\drawline(73,45)(133,45)
\drawline(168,60)(193,85)
\drawline(168,30)(193,5)
\put(3,80){\makebox(0,0)[rb]{ #1 }}
\put(3,0){\makebox(0,0)[rb]{ #3 }}
\put(203,80){\makebox(0,0)[lb]{ #2 }}
\put(203,0){\makebox(0,0)[lb]{ #4 } }
\put(103,50){\makebox(0,0)[b]{ #5 }}
\end{picture}}}

\begin{document}

\TOCauthors{Klaus Lang, Werner R\"uhl}
\articleauthor{Klaus Lang, Werner R\"uhl}
\articleaffil{Fachbereich Physik
Universit\"at Kaiserslautern
67653 Kaiserslautern, Germany
e-mail: lang@gypsy.physik.uni-kl.de
October 1993}
\articletitle{ Critical O($N$) - vector \\
nonlinear sigma - models:\\
A r\'esum\'e of their field\\
structure}
\shortenedtitle{Field structure of critical O($N$) - vector nonlinear $\sigma$
- models}
\begin{abstract}
The classification of quasi - primary fields is outlined. It is proved that the
only
conserved quasi - primary currents are the energy - momentum tensor and the \ON
-
Noether currents. Derivation of all quasi - primary fields and the resolution
of
degeneracy is sketched. Finally the limits $d=2$ and $d=4$ of the space
dimension are
discussed. Whereas the latter is trivial the former is only almost so.\\
(To appear in the Proceedings of the XXII Conference on Differential Geometry
Methods in
Theoretical Physics, Ixtapa, Mexico, September 20-24, 1993)
\end{abstract}

\newcounter{theo}
\section{Some general remarks}
We have studied only a very special example of a critical field theory at
dimensions
$2<d<4$. Nevertheless we believe that the results are relevant for many
critical field
theories, in particular sigma models in a neighbourhood of a free theory. Our
neighbourhood is defined by a \oo expansion.

In this r\'esum\'e we extract results from a series of papers on this subject
[1-7]
published by us in the last three years, and from earlier literature on
conformal field
theory in general \cite{HarmAn,tracts67} or conformal sigma models in
particular
\cite{Va}. These results may
have different status but we condense them equally into "theorems" which should
not be
considered as mathematical theorems but as tested conjectures. General
statements of
quantum field theory and group theory are thus mixed up with conclusions from
low order
perturbative expansions. Let us start with such a theorem which certainly
disappoints many
of the readers:
\Theo{Almost none of the structures of conformal field theory at $d=2$ can be
rediscovered at $2<d<4$.}
\section{Definition of the model}
We start with the partition function
\BEA \label{eq:Z}
{\cal Z} & = & \int D[\vs] D[\al] \exp\bigg\{ -\int dx \Big[
\frac{1}{2}(\partial_\mu\vs)^2 (x)
+ z^{\frac{1}{2}}\al(x)(\vs^2(x)-1)\Big]\bigg\}\qquad
\EEA
where
\BEA
\vs &:& \mbox{\ON - vector, O($d$) - scalar};\nonumber\\
\al &:& \mbox{\ON - O($d$) - scalar};\nonumber\\
d=2\mu &:& \mbox{space - time dimension}\nonumber
\EEA
If $\vs$ and $\al$ are normalized in a standard fashion
\BEA
\EWL S_a(x) S_b(0) \EWR & = & \delta_{ab} \Big( x^2 \Big)^{-\al} \\
\EWL \al(x) \al(0) \EWR & = & \Big( x^2 \Big)^{-\beta}
\EEA
the critical coupling constant $z$ becomes a computable function of $N$, and
\BEQ N\to\infty\,:\qquad z\,=\,\OO
\EEQ
The limit $N\to\infty$ is a free field limit
\BEQ \lim_{N\to\infty}\,\vs(x)\;=\;\vec{s}(x)
\EEQ
but $\vec{s}(x)$ possesses infinitely many components which leads to problems
sometimes. A saddle point expansion of \rf{eq:Z} gives the \oo - expansion.

A critical theory such as this is conformally covariant. Operator product
expansions
(OPE) generate a field algebra ${\cal A}(\vs,\al)$ of the two fundamental
fields $\vs$ and
$\al$
which is associative and possesses a commutation property connected with the
crossing
behaviour of $n$ - point functions. The building blocks of $\cal A(\vs,\al)$
are the conformal or quasiprimary fields (\qps).
\Theo{All \qps belong to representations of the conformal group
characterized by two quantum numbers only: $\delta$, the scaling dimension
under
dilatations and $l$, the tensor rank under space - time rotations.}
These are the elementary representations. In addition the  \qps transform
irreducibly under \ON . We ascribe to them a Young frame $Y$.

Consider the dimension $\delta_{\phi}$ of the \qp $\phi$
\BEA \label{eq:dimPhi}
\delta_{\phi} & = & [\delta_{\phi}] + \eta(\phi) \\
{[}\delta_{\phi}] & : & \mbox{the normal dimension} \nonumber\\
\eta(\phi)=\OO&:& \mbox{the anomalous dimension}
\EEA
By definition
\BEA \label{eq:normaldim}
[\delta_{\phi}] & = & p(\mu-1)+q,\;p,q \in I\!\!N_0 \\
{[}\delta_S] & = & \mu-1
\EEA
So we expect that in the limit $N\to\infty$ $\phi$ tends to a normal product of
$p$
fields $\vs$ with not more than $q$ derivatives (see below).

Each elementary representation $[\delta,l]$ of a \qp possesses a dual
representation $[\delta^\prime,l^\prime]$ (" shadow representation ")
\BEA
\delta^\prime & = & d-\delta+2\,l \mathletter{a}\\
l^\prime & = & l \mathletter{b}
\EEA
The two - point functions
\BEA
\EWL \phi_{[\delta,l]}(x) \phi_{[\delta,l]}(0) \EWR &\quad\mbox{and}\quad&
\EWL \phi_{[\delta^\prime,l^\prime]}(x) \phi_{[\delta^\prime,l^\prime]}(0)
\EWR\nonumber
\EEA
are as kernels and up to a normalization inverse to each other. An $n$ - point
function of $\phi_{[\delta,l]}$ is transformed into an $n$ - point
function of $\phi_{[\delta^\prime,l^\prime]}$ by amputation. Therefore we have
\Theo{The fields $\phi_{[\delta,l]}$ and $\phi_{[\delta^\prime,l^\prime]}$
are dynamically equivalent.}
So from each pair $\phi_{[\delta,l]}$, $\phi_{[\delta^\prime,l^\prime]}$ we
would like
to choose only one representative as basis element of ${\cal A}(\vs,\al)$. We
will in fact be able to do that but in an unexpected fashion.

{}From
\BEA
[\delta^\prime] & = & d- \Big( p (\mu-1) + q\Big) +2l \nonumber\\
& = & (2-p)(\mu-1) + 2 -q +2l
\EEA
we see that the $\al$ - field can be considered as the shadow field of
\BEQ
\Big(\vs^2(x)\Big)_{\mbox{ren.}} \nonumber
\EEQ
since
\BEQ
p=2,\;q=0,\;l=0\;\mbox{implies } [\delta^\prime]=2
\EEQ
Inspection of the action in \rf{eq:Z} also suggests this interpretation of
$\al$.

Next we decompose $q$ in \rf{eq:normaldim} as
\BEQ \label{eq:qtwist}
q\,=\, l+t\,=\,l+2r\qquad \mbox{($t$: twist)}
\EEQ
where $r$ is the number of $\al$ fields bound into $\phi$ at $N\to\infty$ and
$l$
is the number of derivatives. $p$, $l$ and $r$ (or $t$) serve as quantum
numbers in a
neighbourhood of $N\to\infty$.

\section{Classes of qp - fields}

Construction of  the \qps goes by OPE and harmonic analysis. This
automatically orders the \qps according to increasing dimensions $\delta$. From
the interpretation of the quantum numbers $p$, $l$, $t$ in \rf{eq:normaldim},
\rf{eq:qtwist} we can naturally expect these numbers to be bounded by
\BEQ \label{pltlimits}
p\ge 0,\;l\ge0,\;t\ge0
\EEQ
Infact, this is fulfilled by our construction. Most of the shadow fields are
forbidden by \rf{pltlimits} but a few of them are still permitted.\\
We put all \qps with the same $Y$ and $p$ into a class $(Y,p)$. A generic class
looks graphically as Fig.1.
\renewcommand{\bottomfraction}{1}
\renewcommand{\topfraction}{1}
\begin{figure}[t]
\begin{minipage}[t]{2.35in}
\begin{center}
\standardtower
\Caption{A generic class $(Y,p)$}{A generic class $(Y,p)$}
\end{center}
\end{minipage}
\begin{minipage}[t]{2.2in}
\begin{center}
\aStower
\Caption{The class $(\Box,1)$}{The class $(\Box,1)$}
\end{center}
\end{minipage}
\end{figure}
Labels may be multiply occupied by qp -fields, which are distinguished by their
anomalous dimensions (" degeneracy "). Some of the simplest classes look
different
indeed.
\begin{itemize}
\item[(A)] The class $(\YoungS{1},1)$ containing the fundamental field $\vs$.
At $t=0$ there is only the scalar field $\vs$. At the level $t=2$, $l=0$ we
would
expect the shadow field $\vec{S^\prime}$ of $\vs$. But it is not found, this
level
is empty. The level $t=4$, $l=2$ is twofold degenerate.
\item[(B)] The class $(\emptyset,0)$ containing the fundamental field $\al$.
\begin{figure}[b]
\begin{minipage}[b]{\textwidth}
\begin{minipage}[b]{2.3in}
\begin{center}
\atower
\Caption{The class $(\emptyset,0)$}{The class $(\emptyset,0)$}
\end{center}
\end{minipage}
\begin{minipage}[b]{2.3in}
\begin{center}
\Ttower
\Caption{The class $(\emptyset,2)$}{The class  $(\emptyset,2)$}
\end{center}
\end{minipage}
\end{minipage}
\end{figure}
At $t=2$ we have only the $\al$ field (we start counting from $t=2$ in this
case). At
$t=4$ we have only even $l$ and at $t\ge6$, $l=1$ is empty.

Indeed, fusion of two \qps into a third one by OPE
\BEQ
A(x) B(0) \; =\; \Big(x^2\Big)^{\frac{1}{2}(\delta_C-\delta_A-\delta_B)} C(0) +
\ldots
\EEQ
abbreviated as
\BEQ
A \,\otimes \,B\, \rightarrow\,  C
\EEQ
is analogous with the formation of bound states. Two bosonic $\al$'s cannot be
bound
together to a state with odd $l$ and for more than three $\al$'s  $l=1$ is also
excluded by bose symmetrization.
\item[(C)] The class $(\emptyset,2)$ containing the energy - momentum tensor
$T_{\mu\nu}$.

The level $t=0$, $l=0$ has been found unoccupied. The shadow field of $\al$
should
appear on this level, or, according to our remark above, the field
$\Big(\vs^2(x)\Big)_{\mbox{ren.}}$. Thus the sigma - model constraint works and
this field has been eliminated. The energy - momentum tensor field lies at
\BEQ
t\,=\,0,\quad l\,=\,2,\quad\delta\,=\,[\delta]\,=\,2\mu\,=\,d
\EEQ
\end{itemize}
Looking through the classes more carefully, we recognize that the elimination
of
shadow fields has been completed.

In \cite{LR4} we showed that elimination of the shadow field of $\al$ was
directly
related
with a renormalization condition. Using dressed propagators and vertices
(represented
as Polyakov triangles $\bullet$) we have three such conditions
\BEA
\begin{array}[t]{r*5{c}rcll}
\Sprop  &+ &z\SproprenO &+&\ldots & = &  0& \qquad &(S)\Balk{2} \\
\frac{2}{N}\,\aprop &+ &
z\aproprenA &+ &\ldots &  = & 0&\qquad &(\alpha)\Balk{1.5}\\
\Vertex & = &
z\VertexA &+& \ldots &&&&(\Gamma) \Balk{4}
\end{array}\nonumber
\EEA
These three conditions suffice to determine $\eta(S)$, $\eta(\al)$ and $z$.
A generalization of the argument in \cite{LR4} shows validity of
\Theo{The requirement that one (two) shadow field(s) of the fundamental
fields do(es) not show up replaces one (two) renormalization condition(s).}
The status of the proof is still not satisfactory: ${\cal O}(\frac{1}{N^2})$
calculations at best. The theorem (" equivalence theorem ") is very powerful in
practice.

The $\al$ - field produces a field algebra ${\cal A}(\al)$ which is a
subalgebra of
${\cal A}(\vs,\al)$. It contains only \ON - scalars, among them the energy -
momentum
tensor $T_{\mu\nu}$
\BEQ T_{\mu\nu}\,\in\,(\emptyset,2)
\EEQ
Indeed
\BEQ \al\,\otimes\,\al\,\rightarrow\,T
\EEQ
at \OO{}, so p is not conserved at this order. Moreover
\BEQ T\,\otimes\,T\,\rightarrow\,\al
\EEQ
so all ${\cal A}(\al)$ can be generated from $T$ (at $d=2$ T generates not only
{\it Vir} $\times$ $\overline{\mbox{\it Vir}}$ but $W$ algebras as well!).
\Theo{The only conserved qp - currents in ${\cal A}(\vs,\al)$ are
$T_{\mu\nu}$ and $J_{\mu,ab}$, the Noether currents of \ON - symmetry from the
class
$(\YoungA{2},2)$.}
Let us sketch the proof. Denote by $\#Y$ the number of blocks in the Young
frame $Y$.
Then
\BEQ
p\,-\,\#Y\;=\;2n,\quad n\,\in\,I\!\!N_0
\EEQ
This is obvious at $N=\infty$ since $n$ is the number of contractions applied
to the
normal product of $p$ vector fields $\vec{s}$. But in a neighbourhood of
$N=\infty$ it
remains valid due to  standard arguments of harmonic analysis.

Next we use a classical lemma of conformal field theory (\cite{LR5}, Appendix
A) for qp -
fields which are symmetric tensors in spacetime. In fact for $2<d<4$ we have
the
situation of $d=3$: symmetric tensors are sufficient. The lemma says that a qp
-
current is conserved if and only if
\BEQ
l\,\ge\,1,\;\delta\,=\,[\delta]\,=\,2\mu-2+l
\EEQ
i.e.
\BEQ
p\,=\,2,\;l\,\ge\,1,\;t\,=\,0
\EEQ
and
\BEQ
\eta(\phi)\;=\;0
\EEQ
This leaves as candidates the classes
\BEQ
(\YoungS{2},2),\quad(\YoungA{2},2),\quad(\emptyset,2)
\EEQ
In each case the $t=0$ towers are nondegenerate with the following anomalous
dimensions at leading order
\begin{eqnarray*} &
\begin{array}{lcl}
\!\left.\begin{array}[c]{lcl}
(\YoungS{2},2) & : &\ds\frac{\eta(M_l)}{\eta(S)} \\
(\YoungA{2},2) & : & \ds\frac{\eta(J_l)}{\eta(S)}
\end{array}  \right\}
& = & 2\ds\frac{(l-1)(2\mu-2+l)}{(\mu-1+l)(\mu-2+l)} \qquad
\begin{array}[c]{l}
l \quad\mbox{even} \\ l \quad \mbox{odd}
\end{array} \hfill\parbox{1cm}{\BEA\EEA} \\
\begin{array}[c]{lcl}
(\emptyset,2) & : & \ds\frac{\eta(T_l)}{\eta(S)}
\end{array}
& = & \left \{
\begin{array}[c]{l}
0, \qquad (l=2) \\ \\
\!\!\begin{array}[c]{l}
2(l-1)+ \\
\ds\sum\limits_{p=1}^{\frac{1}{2}l-2}
\Big((p+1)!\Big)^2\ds\frac{(2\mu+1+p)_{l-4-2p}}
{(2\mu+1)_{l-4}},\\
\hfill (l\ge 4, \mbox{even}) \hfill
\end{array}
\end{array}
\right.
\label{eq:etaT}\parbox{1cm}{\BEA\EEA}
\end{array} &
\end{eqnarray*}
The curves for the expression \rf{eq:etaT} are presented in \cite{LR6}, Fig. 6.
None of these
functions changes sign. They vanish identically for $J_1$ and $T_2$ and are
otherwise
different from zero for all $2<d<4$. It is also important to guarantee that no
empty
levels are filled up at higher orders of $\frac{1}{N}$ or that degeneracy
appears this
way. The first is made sure by crossing symmetry, the second possibility can at
present
not be excluded.

\section{Fusion}
Each \qp has a pedigree of fusion.
\begin{figure}[t]
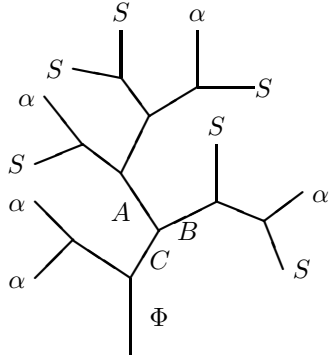

\begin{center}
\fusetree
\end{center}
\caption[A pedigree of fusion]{A pedigree of fusion}
\end{figure}
The internal lines are arbitrary qp - fields which can be produced from the
parents.
This means that the fusion coefficients effective at a vertex must be nonzero:
\BEQ
f_{AB}^C\;\neq\;0
\EEQ
\Theo{Fusion coefficients vanish only if the corresponding Littlewood -
Richardson coefficients of \ON{}  are zero or if this follows from a crossing
symmmetry
selection rule.}
As an example let $A=B$ scalar and the Littlewood - Richardson coefficient be
symmetric (antisymmetric) under exchange of A and B. Then odd (even) $l$ are
forbidden
for $C$. Another example is the fusion
\BEQ
\al\,\otimes\,\vs
\EEQ
which leads to any level of the class $(\YoungS{1},1)$ at \OO already, but in
the
class of degenerate levels only to one linear combination of \qps. So to
resolve degeneracies we have to consider different pedigrees with the same
final level.
\Theo{The \qps with $l=0$ are never degenerate.}
This corresponds to the uniqueness of a ground state in QM.

We introduce the concept of " dominant channel fusion " (DCF). This kind of
fusion
acts already at ${\cal O}(1)$ and produces scalar \qps of the type
\BEQ \label{eq:longlabel}
(Y,p;[\delta],l)\;=\; (\YoungSg{p},p;p(\mu-1)+2r,0)
\EEQ
from \qps of the same type. Let two such \qps with labels $\{p_1,r_1\}$,
$\{p_2,r_2\}$ be given. The resulting field has labels $\{P,R\}$ with
\BEA
P& = & p_1+p_2\\
R &= & r_1+r_2
\EEA
For DCF normal dimensions are additive and degeneracy does not occur. Only
symmetric
\ON{} tensors are produced by definition. Pedigrees with DCF at each vertex
produce a
\qp of type \rf{eq:longlabel} which depends only on the numbers $p$ of \vs{}
fields
and $r$ of \al{} fields entering and not on the form of the pedigree. In other
words:
DCF is abelian.

We denote the \qps \rf{eq:longlabel} by $M^{\{p,r\}}_0$. Any \qp on the level
\BEQ
(\YoungSg{p},p;p(\mu-1)+2r+l,l) \nonumber
\EEQ
is denoted $M^{\{p,r\}}_{l,k}$ where $k$ is introduced to take account of the
degeneracy. We are interested in the fusion process
\BEQ \label{eq:DCF}
M^{\{p_1,r_1\}}_0\,\otimes\,M^{\{p_2,r_2\}}_0\,\rightarrow\,
M^{\{p_1+p_2,r_1+r_2\}}_{l,k}
\EEQ
If we keep
\BEQ
P\;=\;p_1+p_2,\qquad R\;=\;r_1+r_2
\EEQ
fixed but let $p_1$, $r_1$ run, we obtain different combinations of
$M^{\{P,R\}}_{l,k}$
which can be resolved.

Technically one considers the four - point functions
\BEQ \label{eq:fourpoint}
\EWL M^{\{p_1,r_1\}}_0(y_1)\,M^{\{p_2,r_2\}}_0(y_2)\,
M^{\{p_1^\prime,r_1^\prime\}}_0(y_3)\,M^{\{p_2^\prime,r_2^\prime\}}_0(y_4) \EWR
\EEQ
with fixed
\BEA
P &=\quad p_1+p_2 \quad = & p_1^\prime+p_2^\prime \nonumber\\
R &=\quad r_1+r_2 \quad = & r_1^\prime+r_2^\prime
\EEA
On the one hand these four - point functions \rf{eq:fourpoint} are calculated
from a
$2(P+R)$ - point function involving $2P$ \vs{} fields and $2R$ \al{} fields by
OPE
reduction via DCF.
This is mainly a combinatorical task bringing in the " replica parameters "
$p_1$, $r_1$, $p_2$, $r_2$, $p_1^\prime$, $r_1^\prime$, $p_2^\prime$,
$r_2^\prime$ and,
at \OO, the connected four - point functions
\BEQ
\EWL SSSS \EWR_{\mbox{conn}},\qquad \EWL \al\al\al\al \EWR_{\mbox{conn}},\qquad
\EWL
 \al S\al S\EWR_{\mbox{conn}} \nonumber
\EEQ
which are explicitly known [3,4,5]. Crossing between the unprimed factors
exchanges
\BEQ
p_1\,\leftrightarrow\,p_2,\qquad r_1\,\leftrightarrow\,r_2
\nonumber\EEQ
so that we can use the crossing symmetric combinations
\BEQ
t_1\;=\;r_1r_2,\quad t_2\;=\;p_1p_2,\quad t_3\;=\;p_1 r_2+p_2 r_1
\EEQ

On the other hand we compare the four - point function \rf{eq:fourpoint} with
conformal exchange amplitudes (this is an element of harmonic analysis).
\renewcommand{\bottomfraction}{1}
\renewcommand{\topfraction}{1}
\begin{figure}
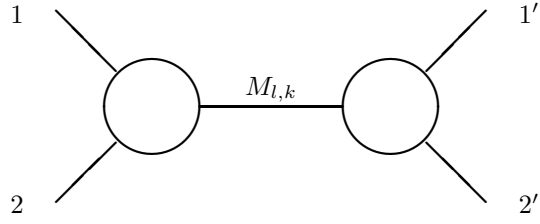

\begin{center}
\amplitude{1}{$1^\prime$}{2}{$2^\prime$}{$M_{l,k}$}
\vspace{3em}
\end{center}
\caption[Conformal exchange amplitude]{Conformal exchange amplitude}
\end{figure}
This allows us to extract expressions for
\BEQ \label{eq:fuse}
\sum\limits_{k} \,f_{12}^{M_{l,k}}\,f_{1^\prime2^\prime}^{M_{l,k}}
\EEQ
and
\BEQ \label{eq:fuseeta}
\sum\limits_{k} \,f_{12}^{M_{l,k}}\,f_{1^\prime2^\prime}^{M_{l,k}}
\eta(M_{l,k})
\EEQ
The fusion constants $f_{12}^{M_{l,k}}$ are functions of the replica parameters
\BEA
f_{12}^{M_{l,k}} & =& F_{l,k}(p_1,r_1;p_2,r_2)
\EEA
By a simultanous diagonalization procedure for the two expressions obtained for
\rf{eq:fuse}, \rf{eq:fuseeta} we can extract the fusion coefficients and the
anomalous
dimensions . The fusion coefficients are obtained in the form
\BEA \label{eq:fM}
f_{12}^{M_{l,k}} & = & \mbox{\begin{minipage}[t]{8cm}
polynomial in the replica parameters giving $(-1)^l$ \mbox{under} crossing
times an algebraic
function depending homo\-genously on $t_1$, $t_2$, $t_3$ \end{minipage}}
\EEA
We have in fact solved the following cases \cite{LR7}

\quad\begin{tabular}[t]{ll}
 $P=0$, $R$ arbitary $>0$:&
levels $0\le l\le 6$ and $t=2R$ in the class
$(\emptyset,0)$. \\&Degeneracy sets in at $R\ge 4$ and $ l\ge 4$,\\
$R=0$, $P$ arbitary $>0$: &
levels $0\le l\le 6$ and $t=0$ in the classes \\ &
$(\YoungSg{P},P)$. Degeneracy sets in at \\& $P\ge 4$ and $ l\ge 4$.
\end{tabular}

\noindent In both cases the anomalous dimensions are
\BEA
\frac{\eta(M_{l,k})}{\eta(S)} & = & \mbox{\begin{minipage}[t]{8cm} rational
functions
of $\mu$ at leading order. \end{minipage}}
\EEA
and the algebraic function in \rf{eq:fM} reduces to a (nonhomogeneous)
polynomial of
either $t_1$ or $t_2$.

If $R\,P \neq0$, degeneracy starts already at $R+P\ge 3$, $l\ge 2$. We resolved
only
the cases $0 \le l \le 3$. Moreover we find
\BEA
\frac{\eta(M_{l,k})}{\eta(S)} & = & \mbox{\begin{minipage}[t]{8cm} algebraic
(irrational) function of $\mu$ at leading order. \end{minipage}}
\EEA
Many infinite sequences of anomalous dimensions are known now and in these
sequences
we can study limits. Consider a tower of nondegenerate qp - fields
$M^{\{P,R\}}_l$, $P$, $R$
fixed, $l$ running. Then
in the DCF process \rf{eq:DCF} the pair of \qps on the left hand side is
uniquely
determined. At leading order in $\frac{1}{N}$ we find
\BEA
\lim_{l\to\infty} \eta(M^{\{P,R\}}_l)&=&
\eta(M^{\{p_1,r_1\}}_0)+\eta(M^{\{p_2,r_2\}}_0)
\EEA
Instead in the case of degeneracy
\BEA
\eta(M^{\{P,R\}}_l) & = & {\cal O}\Big(\frac{l^2}{N}\Big)
\EEA
which makes the $\frac{1}{N}$ expansion asymptotic only if $N\gg l^2$. We could
also
think of keeping $l$ fixed and letting $P$, $R$ run. Then
\BEA
\eta(M^{\{P,R\}}_l) & = & {\cal
O}\Big(\frac{1}{N}\times \mbox{second order polynomial in
$P$ and $R$}\Big)
\EEA
imposing a similiar restriction on $N$.

We emphasize that our method of constructing the states $M^{\{P,R\}}_l$ by
forcing all internal \qps of the pedigree to have tensor rank zero may be too
restrictive for large $l$. In a forthcoming article we will study an
alternative
algorithm which remains correct at large $l$ as well.

\section{The limits $d\searrow 2$ and $d\nearrow 4$}
For any $2<d<4$ the limit $N\to\infty$ leads to a free field theory. In this
limit
each \qp $\phi \in {\cal A}(\vs ,\al)$ possesses a corresponding \qp $\varphi$
in the
free field algebra ${\cal A}_0(\vec{s})$. In Green functions involving \al{}
fields we
may first amputate them and perform the limit afterwards. At the boundaries
$d=2$, $d=4$
the behaviour of coupling constant and critical indices
\BEA
\eta(\phi) &=& \sum\limits_{k=1}^\infty\,\ds\frac{\eta_k(\phi)}{N^k}\\
\label{eq:etaS}
\eta_1(S) &=& 2\,\frac{\sin\pi\mu}{\pi}\,\frac{\Gamma(2\mu-2)}
{\Gamma(\mu+1)\Gamma(\mu-2)}\\
z&=& \sum\limits_{k=1}^\infty\,\ds\frac{z_k}{N^k}
\EEA
concerning their zero orders in $d$ is listed in table \ref{tb:zeros}.
\begin{table}
\begin{center}
\[\begin{array}[t]{|l|c|c|}
\hline
& d=2 & d=4\\ \hline
z_1 &0&2\\
z_2 &0&2\\
\eta_1(S) & 1&2\\
\eta_2(S) & 1&2\\
\eta_1(\phi),\,\phi\neq S & 1 & 1\\
\hline
\end{array}\]
\end{center}
\caption[Table of zero orders]{Table of zero orders.}
\label{tb:zeros}
\end{table}

All critical exponents vanish at both limits. These limits are therefore
connected
with free field theory.

At $d=4$ we obtain a free field theory in the trivial sense that
\BEA
\lim_{d\nearrow 4} \vs(x) & = & \vec{s}(x),\quad \triangle
\vec{s}(x)\;=\;0\nonumber\\
\vec{s}(x) &:& \mbox{$N$ - component \ON{} - vector field}
\EEA
As a test we can calculate the limit of $\EWL \al \vs \al \vs\EWR$ after
amputation.
This limit $d=4$ is assumed fieldwise and is an isomorphism of field algebras
in
the straightforward sense. Let $A,B,C\in{\cal A}(\vs, \al)$
\BEA
A(x)\,B(0) &= & \Big(x^2\Big)^{\frac{1}{2}(\delta_C-\delta_A-\delta_B)(\mu)}
f_{AB}^C(\mu) \,C(0)\,+\ldots
\EEA
Then if $a$, $b$, $c$ are the corresponding free fields
\BEA
a(x)\,b(0) & = & \Big(x^2\Big)^{\frac{1}{2}(\delta_C-\delta_A-\delta_B)(2)}
f_{AB}^C(2) \,c(0)\,+\ldots
\EEA
The limit $\mu\to 2$ is performed termwise.

This is not true at the other limit $d=2$. First we consider the two conserved
qp -
currents
\BEA
\phi &:& T_{\mu\nu}\quad\mbox{or}\quad J_{\mu,ab} \nonumber
\EEA
which have well defined local field limits
\BEA
\varphi&:& t_{\mu\nu}\quad\mbox{or}\quad j_{\mu,ab} \nonumber
\EEA
Both $T$ and $J$ can be constructed from fusion of $\vs\otimes\vs$. We
introduce the
Ward identities in any ad hoc normalization and normalize the fields $\phi\ni
\{T,J\}$, $\varphi\ni \{t,j\}$ relative to the same Ward identities. Conformal
invariance implies the same scaling dimension and tensor structure for $\phi$
and
$\varphi$ so that
\BEA
\EWL \phi(x)\,\phi(0)\EWR &=& C_\phi(\mu)\,\EWL \varphi(x)\,\varphi(0)\EWR
\EEA
By explicit calculation we find
\BEA
\lim_{\mu \searrow 1}\,C_\phi(\mu)&=&1-\ds\frac{l+1}{N}+{\cal O}
\Big(\frac{1}{N^2}\Big)\\
l&=& \mbox{tensor degree of $\phi$ (1 or 2)}\nonumber
\EEA

Ward identities can be derived from the two - point functions. Instead of
normalizing
fields by three - point functions and comparing the two - point functions we
can
introduce a standard normalization of two - point functions
\BEA
\EWL \phi(x)\,\phi(0)\EWR &=& \Big(x^2\Big)^{-\delta_\phi}\cdot
\mbox{tensor}(x)
\EEA
with the tensor factors connected to Gegenbauer polynomials which can be
submitted to
an ad hoc normalization, say $C_l^{\mu-1}(1)=1$, too. Doing that, the factors
$C_\phi (\mu)$ appear in the three - point functions as fusion coefficients. It
becomes clear that the appearance of such factors is quite general. Consider
the
fusion of $n$ fields \vs{} by DCF into the field $M_0^{\{n,0\}}$. In the free
field
limit this corresponds to taking the Wick normal product
\BEQ
:\vec{s}^{\;n}_\otimes:\nonumber
\EEQ
Two such fields multiply as
\BEA
:\vec{s}^{\;p_1}_\otimes:(x)\;:\vec{s}^{\;p_2}_\otimes:(0) & = &
:\vec{s}^{\;p_1+p_2}_\otimes:(0)\,+\ldots
\EEA
whereas DCF yields
\BEA
M_0^{\{p_1,0\}}(x)\,M_0^{\{p_2,0\}}(0)&=& f\,\Big(x^2\Big)^{{\cal O}(\mu-1)}\,
M_0^{\{p_1+p_2,0\}}(0) \,+\ldots
\EEA
The exponent of $x^2$ contains only anomalous dimensions and tends to zero at
$\mu=1$.
Computation of $f$ gives
\BEA
f(\mu)&=& 1+ \ds\frac{\mu}{(\mu-1)(\mu-2)}\eta(S)\,p_1 p_2+
{\cal O}\Big(\frac{1}{N^2}\Big)
\EEA
so that, with \rf{eq:etaS}
\BEA
\lim_{\mu\searrow 1}\,f(\mu) &= & 1 -\ds\frac{p_1 p_2}{N} +
{\cal O}\Big(\frac{1}{N^2}\Big)
\EEA
Then we end up with a final
\Theo{The $d=2$ limit is into the \underline{universality class} of the
polynomial algebra of free fields. Fusion coefficients are \OO deformed with
respect
to free field theory.}
In particular this implies that exponential expressions of free fields ("
vertex
operators ") cannot arise. Moreover the $\epsilon=d-2$ expansions (which are in
the
literature since about 1976) are correct only if applied to critical indices
and not
to amplitudes. To our knowledge this restriction has never been clearly
expressed
before.

\end{document}